# Inter-rater reliability and convergent validity of F1000Prime peer review

Lutz Bornmann

Division for Science and Innovation Studies

Administrative Headquarters of the Max Planck Society

Hofgartenstr. 8,

80539 Munich, Germany.

Email: bornmann@gv.mpg.de

**Abstract**

Peer review is the backbone of modern science. F1000Prime is a post-publication peer review system of the biomedical literature (papers from medical and biological journals). This study is concerned with the inter-rater reliability and convergent validity of the peer recommendations formulated in the F1000Prime peer review system. The study is based on around 100,000 papers with recommendations from Faculty members. Even if intersubjectivity plays a fundamental role in science, the analyses of the reliability of the F1000Prime peer review system show a rather low level of agreement between Faculty members. This result is in agreement with most other studies which have been published on the journal peer review system. Logistic regression models are used to investigate the convergent validity of the F1000Prime peer review system. As the results show, the proportion of highly cited papers among those selected by the Faculty members is significantly higher than expected. In addition, better recommendation scores are also connected with better performance of the papers.

**Keywords**

F1000; F1000Prime; Peer review; Highly cited papers; Adjusted predictions; Adjusted predictions at representative values

# 1 Introduction

The first known cases of peer-review in science were undertaken in 1665 for the journal *Philosophical Transactions of the Royal Society* (Bornmann, 2011). Today, peer review is the backbone of science (Benda & Engels, 2011); without a functioning and generally accepted evaluation instrument, the significance of research could hardly be evaluated. Callaham and McCulloch (2011) therefore regard peer review as an important advance in scientific progress. Peers or colleagues asked to evaluate manuscripts in a peer review process take on the responsibility for ensuring high standards in their disciplines. Although peers active in the same field might be unaware of other perspectives, they “are said to be in the best position to know whether quality standards have been met and a contribution to knowledge made” (Eisenhart, 2002, p. 241). Peer evaluation thus entails a process by which a jury of equals active in a given scientific field convenes to evaluate scientific outcomes. Examples from the field of the physical sciences of what may happen in absence of peer review even in a reputable venue like the arXiv preprint server (http://arxiv.org) have been recently discussed in Iorio (2014).

According to Marsh, Bond, and Jayasinghe (2007) “the quantitative social science research tools used by psychologists (with their focus on reliability, validity, and bias) are uniquely appropriate to evaluate the peer review process” (p. 33). As an assessment tool, peer review is asked to be reliable, fair, and valid. However, critics of peer review argue that: (1) reviewers rarely agree on whether to recommend that a manuscript be published, thus making for poor reliability of the peer review process; (2) reviewers’ recommendations are frequently biased, that is, judgments are not based solely on scientific merit, but are also influenced by personal attributes of the authors or the reviewers themselves (where the fairness of the process is not a given); (3) the process lacks (predictive) validity because there is little or no relationship between the reviewers’ judgments and the subsequent usefulness of the work to the scientific community, as indicated by the frequency of citations of the work in later scientific papers (Bornmann, 2011).

In January 2002, a new type of peer-review system has been launched (in the field of the medical and biological sciences), when around 5000 Faculty members were asked “to identify, evaluate and comment on the most interesting papers they read for themselves each month – regardless of the journal in which they appear” (Wets, Weedon, & Velterop, 2003, p. 251). The so-called F1000Prime[1] peer review system is considered not as an ex-ante appraisal of manuscripts offered to a journal for publication, but an ex-post appraisal of papers already published in journals.[2] Even if the recommendations for F1000Prime (“good,” “very good,” or “exceptional”) are produced by peers (here: Faculty members) after publication has taken place, the question still arises of the quality of these recommendations: Are they reliable, fair and valid?

A number of studies have already been conducted into F1000Prime recommendations (Bornmann & Leydesdorff, 2013; Li & Thelwall, 2012; Mohammadi & Thelwall, 2013; Waltman & Costas, 2014; Wardle, 2010). Waltman and Costas (2014) have published the most comprehensive analysis so far, based on more than 100,000 publications which were rated in F1000Prime. All these studies have essentially concerned the connection between the recommendations for the publications and their citation impact. However these studies have either used the total score for each publication (e.g., Bornmann & Leydesdorff, 2013), which is derived from the separate ratings for a publication, or they have included the best score for a publication in their analysis (Waltman & Costas, 2014). The validity of the individual recommendations have not yet been tested. Since the evaluation of a peer review system reflects not only the validity but also the reliability, this study also tests the agreement between the recommendations of a publication. Fairness – the third measure of goodness for professional evaluations – is not addressed in the current study. A separate comprehensive study would be needed for an analysis of the F1000Prime peer review system, to investigate possible biases related to characteristics of the authors, Faculty members and publications.

[1] The name of the Faculty of 1000 (F1000) service post-merging is F1000Prime. F1000 is the name of the umbrella company, which has three distinct services: F1000Prime, F1000Posters, and F1000Research.
[2] With ScienceOpen (https://www.scienceopen.com) a similar post-publication peer-review is going to be launched, which is not limited to medicine and biology.

# 2 Peer ratings provided by F1000Prime

F1000Prime is a post-publication peer review system of the biomedical literature (papers from medical and biological journals). This service is part of the Science Navigation Group, a group of independent companies that publish and develop information services for the professional biomedical community and the consumer market. F1000 Biology was launched in 2002 and F1000 Medicine in 2006. The two services were merged in 2009 and today form the F1000Prime database. Papers for F1000Prime are selected by a peer-nominated global "Faculty" of leading scientists and clinicians who then rate them and explain their importance (F1000, 2012). This means that only a restricted set of papers from the medical and biological journals covered is reviewed (Kreiman & Maunsell, 2011; Wouters & Costas, 2012).

The Faculty nowadays numbers more than 5,000 experts worldwide, assisted by 5,000 associates, which are organized into more than 40 subjects (which are further subdivided into over 300 sections). On average, 1,500 new recommendations are contributed by the Faculty each month (F1000, 2012). Faculty members can choose and evaluate any paper that interests them; however, "the great majority pick papers published within the past month, including advance online papers, meaning that users can be made aware of important papers rapidly" (Wets, et al., 2003, p. 254). Although many papers published in popular and high-profile journals (e.g. *Nature*, *New England Journal of Medicine*, *Science*) are evaluated, 85% of the papers selected come from specialized or less well-known journals (Wouters & Costas, 2012). "Less than 18 months since Faculty of 1000 was launched, the reaction from scientists has been such that two-thirds of top institutions worldwide already subscribe, and it was the recipient of the Association of Learned and Professional Society Publishers (ALPSP) award for Publishing Innovation in 2002 (http://www.alpsp.org/about.htm)" (Wets, et al., 2003, p. 249). The F1000Prime database is regarded as a meaningful aid to scientists seeking indications as to the most relevant papers in their subject: "The aim of Faculty of 1000 is not to provide an evaluation for all papers, as this would simply exacerbate the 'noise', but to take advantage of electronic developments to create the optimal human filter for effectively reducing the noise" (Wets, et al., 2003, p. 253).

The papers selected for F1000Prime are rated by the members as "Good," "Very good" or "Exceptional" which is equivalent to scores of 1, 2, or 3, respectively.[3] In many cases a paper is not evaluated by one member alone but by several members. The FFa (F1000Prime Article Factor), which is given as a total score in the F1000Prime database, is calculated from the different recommendations for one publication. Besides the recommendations, Faculty members also tag the publications with classifications, such as:

- Changes Clinical Practice: the paper recommends a complete, specific and immediate change in practice by clinicians for a defined group of patients.
- Confirmation: validates previously published data or hypotheses.
- Controversial: challenges established dogma.
- Good for Teaching: key paper in field and/or well written.
- Interesting Hypothesis: presents new model.
- New Finding: presents original data, models or hypotheses.
- Novel Drug Target: suggests new targets for drug discovery.
- Refutation: disproves previously published data or hypotheses.

[3] According to the F1000 Outreach Director, Iain Hrynaszkiewicz, Faculty members have been polled about publishing negative or critical evaluations, as a core part of the service, but they felt allowing negative recommendations would damage the service as F1000Prime was designed to identify only the best published research in biology and medicine. Faculty Members are able to include constructive criticism in their recommendations of papers, but the overall tone has to be positive for the recommendations to meet the F1000Prime publication criteria. If a paper is recommended that another member feels is flawed, they can publish a dissenting opinion to explain their reasoning. This also helps ensure the service is unbiased. Users (often the articles' authors) are also free to make user comments.

- Technical Advance: introduces a new practical/theoretical technique, or novel use of an existing technique.

The classifications, recommendations and bibliographic information for publications form the fully searchable F1000Prime database containing more than 100,000 records (End of 2013). Seen as a whole, the F1000Prime database is not regarded solely as an aid for scientists seeking the most relevant papers in their subject area, but also as an important tool for research evaluation purposes. Thus, for example, Wouters and Costas (2012) write that "the data and indicators provided by F1000Prime are without doubt rich and valuable, and the tool has a strong potential for research evaluation, being in fact a good complement to alternative metrics for research assessments at different levels (papers, individuals, journals, etc.)" (p. 14).

# 3 Methods

## 3.1 Construction of the data set to which bibliometric data and indicator are appended

In January 2014, F1000Prime provided me with data on all recommendations (and classifications) made and the bibliographic information for the corresponding papers in their system (n=149,227 records). The dataset contains a total of 104,633 different DOIs which, with a few exceptions, are all individual papers (not all DOIs refer to a specific paper). Research for the present paper extracting a range of bibliometric data and indicators for each paper from an in-house database at the MPG administered by the Max Planck Digital Library (MPDL) and based on the Web of Science (WoS, Thomson Reuters). In order to be able to establish a link between the individual papers and the bibliometric data/indicators, two procedures were followed in this study: A total of 90,436 papers in the dataset could be matched with a paper in the in-house database using the DOI. (2) For 4,205 of the 14,197 remaining papers, no match was possible with the DOI, but could be achieved with the name of the first author, the journal, the volume and the issue. Bibliometric data/indicators were then available for 94,641 papers of the 104,633 in total (91%). This percentage approximates to the 93% quoted by Waltman and Costas (2014), who used a similar procedure to match data from F1000Prime with bibliometric data in their own in-house database.

### 3.2 Dataset for the evaluation of reliability and validity

From the total of 149,227 records in the F1000Prime dataset, 317 cannot be included in this study because they contain no recommendation between 1 and 3. Of the remaining records, 2,128 cannot be used in the analysis because they cannot be uniquely (with the PubMed ID) associated with a particular publication.[4] Overall, this leaves 146,782 records available for the assessment of reliability. For the evaluation of validity, the DOI is used to match these records with the dataset which contains the bibliometric data/indicators from the in-house database (see description of dataset generation in section 3.1). Of the total of 146,782 records a match could be made for 121,895 (83%).

In order to determine the citation impact of papers reliably, a sufficiently wide citation window of at least three years should be available for the papers (Bornmann & Daniel, 2008a). This means that papers with F1000Prime recommendations from recent publication years (after 2011) cannot be included in the citation analyses. In addition, the in-house database used in this study does not include citation impact values for all papers (e.g. not for all document types). The validity analyses involve a total of 96,202 records (66% of 146,782 records) or 95,325 records (65%), if the Journal Impact Factor (JIF) is included in the analysis. Since the probability of citations also depends on the reputation of the journal in which a paper appears (Bornmann, Mutz, Marx, Schier, & Daniel, 2011), the JIF is checked in the statistical analysis. The JIF is the average number of times papers from a journal published in the past two years have been cited in the current year. For example, a 2009 JIF of

[4] The analysis only covers papers which can also be found in PubMed.

4.25 means that, on average, a paper published in the journal in 2007 or 2008 received 4.25 citations in 2009. The JIF used for a publication in this study is that for the publication year (and not the most current JIF available).

### 3.3 Indicators for the measurement of citation impact

Cross-field and cross-time-period comparisons of citation impact are only possible if the impact is normalized (standardized) (Bornmann & Marx, 2013; Schubert & Braun, 1986). For its citation impact to be normalized, a paper needs to have a reference set: all the papers published in the same publication year and subject category (and document type). Percentiles have been proposed as an alternative to normalization on the basis of central tendency statistics (arithmetic averages of citation counts) (Bornmann, Leydesdorff, & Mutz, 2013; Bornmann & Mutz, 2011; Bornmann, Mutz, Marx, et al., 2011; Schreiber, in press). Percentiles are based on an ordered set of publications in a reference set, whereby the fraction of papers at or below the citation counts of a paper in question is used as a standardised value for the relative citation impact of this focal paper. This value can be used for cross-field and cross-time-period comparisons. If the normalized citation impact for more than one paper is needed in a research evaluation study, this percentile calculation is repeated (by using corresponding reference sets for each one).

In this study, the percentile indicator $P_{top\ 10\%}$ is used to measure the citation impact of papers. $P_{top\ 10\%}$ is a binary variable with the value 1 if a paper belongs to the top 10% most frequently cited publications (otherwise the value is 0). A paper belongs to the top 10% most frequently cited if it is cited more frequently than 90% of the papers published in the same subject category and in the same publication year (and as the same document type). $PP_{top\ 10\%}$ – that is the proportion of the top 10% most frequently cited papers – is one of the methods most often used to determine scientific excellence (Bornmann, in press). This indicator is also used in the SCImago Institutions Ranking (Bornmann, de Moya Anegón, & Leydesdorff, 2012) and is seen as the most important indicator in the Leiden Ranking of institutions worldwide by the Centre for Science and Technology Studies (Leiden University, The Netherlands) (Waltman et al., 2012). As early as the late 1980s, Evered, Hamett, and Narin (1989) used the indicator to investigate the impact of different modes of research funding.

$P_{top\ 10\%}$ was generated with the help of percentiles for every paper in the data set of this study. The percentiles were calculated based on a method used by InCites (Thomson Reuters, Bornmann, et al., 2013).

### 3.4 Sample and population

Williams and Bornmann (2014) argue that, even though all records have been collected for an institution, a research group, or – here – all papers in the F1000Prime database, the use of inferential statistics and significance testing is both common and desirable. It could be argued that there is no need to compute significance tests or confidence intervals (CIs) given bibliometric population data. That is, we do not need to estimate parameters or make inferences about the larger population because the information on the entire population of papers is available. Two rationales are typically offered for treating what appears to be a population as though it were a sample. First, the current cases might be thought of as being a sample from a larger super population that includes future cases as well (Gelman, 2009). A second rationale, and a perhaps more compelling one, is to think of observed cases as repeated trials that are products of an underlying stochastic process.

For bibliometrics, Williams and Bornmann (2014) argue that the observed citation impact of papers (measured by percentiles or $P_{top\ 10\%}$) allows us to make inferences about the underlying process that generated those impacts and the extent to which citations may have been influenced by random factors. The success of a paper is presumably affected by the quality of the research reported in the papers, but is also partly determined by chance. As is shown in the overview Bornmann and Daniel (2008b), there are a range of factors – besides the quality of the publication – which influence citation impact. Thus the impact may be influenced by the number of authors, the number of pages in a publication or the language in which a paper is written. The reputation of the authors can also play a role.

Overall around 100,000 records are available for the assessment of reliability and validity in this study. With such a large dataset, cross-validating the results of the statistical analysis seem advisable. In connection with the results of regression models, Sheskin (2007) recommends the following procedure: „It cannot be emphasized too strongly that upon conducting a regression analysis, it is recommended that the resulting regression model be cross-validated. By cross-validating a model, a researcher can demonstrate that it generates consistent results, and will thus be of theoretical or of practical value in making predictions among members of the reference population upon which the model is based" (p. 1240). With one of these methods which could be used for cross-validation, the results of a statistical analysis are replicated on two random samples, each containing half of the original sample. Since the statistical evaluations for reliability involve the use of subgroups containing highly reduced numbers of cases, only the statistical analysis of validity is cross-validated.[5]

# 4 Results

## 4.1 Assessments of reliability

In everyday life, "intersubjectivity is equated with realism" (Ziman, 2000, p. 106). Therefore, scientific discourse is also distinguished by its striving for consensus. Scientific activity would clearly be impossible unless scientists could come to similar conclusions. According to Wiley (2008) "just as results from lab experiments provide clues to an underlying biological process, reviewer comments are also clues to an underlying reality (they did not like your grant for some reason). For example, if all reviewers mention the same point, then it is a good bet that it is important and real" (p. 31).

[5] The statistical software package Stata 13.1 (http://www.stata.com/) is used for this study; in particular, we make use of the Stata commands icc, kappa, logit, margins, marginsplot, and meta.

Cicchetti (1991) defines inter-rater reliability "as the extent to which two or more independent reviews of the same scientific document agree" (p. 120). Manuscripts are rated reliably when there is a high level of agreement between independent reviewers. In many studies of peer review the intraclass correlation coefficient measures the extent of agreement within peer review groups (Mutz, Bornmann, & Daniel, 2012). "The intraclass correlation coefficient (ICC) … is a variance decomposition method to assess the portion of overall variance attributable to between-subject variability. … Raters are assumed to share common metric and homogeneous variance (i.e., intraclass variance)" (von Eye & Mun, 2005, p. 116). The ICC can vary between -1.0 and +1.0. However, high agreement alone with low between-reviewer variability cannot result in high reliability because a certain level of agreement can be expected to occur on the basis of chance alone. Therefore the Kappa coefficient figures in many studies on peer review as a measure of between-reviewer variability.

Kappa (k) statistically indicates the level of agreement between two or more raters. If the raters are in complete agreement then k = 1; if k is near 0, the observed level of agreement is not much higher than by chance (von Eye & Mun, 2005). If a manuscript meets scientific standards and contributes to the advancement of science, it can be expected that two or more reviewers will agree on its value. A meta-analysis by Bornmann, Mutz, and Daniel (2011) of 48 studies on the reliability of agreement between reviewers' ratings in journal peer review reports overall agreement coefficients of mean $ICC/r^2$=.34 and mean k=.17. According to Fleiss's (1981) guidelines, k coefficients between 0 and 0.2 indicate a slight level of reviewer agreement. Cicchetti and Sparrow (1981) interpret ICC<.4 as a low level of inter-rater reliability. An ICC of .4 means that on average two ratings of the same manuscript are correlated with r=.4 or that 40% of the total variance of the ratings is explained by the manuscripts.

Table 1 shows the distribution of the different numbers of recommendations which Faculty members expressed for a paper. Overall the papers received between one and 20 recommendations from different Faculty members. Most of the papers (around 94%) have one recommendation (around 81%) or two recommendations (around 13%) by Faculty members. As an example, Table 2 is a cross-classification of the individual assessments for the papers with two recommendations by Faculty members. The numbers in the main diagonal refer to those papers where the assessments of the two Faculty members agree. As the figures show, the recommendations agree for 7,357 papers (4,315 + 2,768 + 274) of the total of 14,476 papers (around 51%). For 784 Papers (391 + 393, around 5%), the recommendations differ significantly with scores of 1 and 3.

Table 1.

Distribution of the different numbers of recommendations expressed by Faculty members for a paper

| Number of recommendations | Frequency | Percent | Cumulative percent |
|---|---|---|---|
| 1 | 91,467 | 80.96 | 80.96 |
| 2 | 14,476 | 12.81 | 93.78 |
| 3 | 4,225 | 3.74 | 97.52 |
| 4 | 1,602 | 1.42 | 98.93 |
| 5 | 637 | 0.56 | 99.50 |
| 6 | 266 | 0.24 | 99.73 |
| 7 | 138 | 0.12 | 99.85 |
| 8 | 72 | 0.06 | 99.92 |
| 9 | 49 | 0.04 | 99.96 |
| 10 | 17 | 0.02 | 99.98 |
| 11 | 6 | 0.01 | 99.98 |
| 12 | 7 | 0.01 | 99.99 |
| 13 | 2 | | 99.99 |
| 14 | 5 | | 99.99 |
| 15 | 2 | | 100.00 |
| 16 | 1 | | 100.00 |
| 17 | 2 | | 100.00 |
| 20 | 1 | | 100.00 |

Table 2.

Distribution of recommendations for a paper with two Faculty members

| | Faculty member 1 | | | | |
|---|---|---|---|---|---|
| Faculty member 2 | Recommen-dation | 1 | 2 | 3 | Total |
| | 1 | 4,315 | 2,551 | 393 | 7,259 |
| | 2 | 2,615 | 2,768 | 565 | 5,948 |
| | 3 | 391 | 604 | 274 | 1,269 |
| | Total | 7,321 | 5,923 | 1,232 | 14,476 |

ICC and k were calculated for the judgement of two to nine Faculty members per paper. No coefficients were calculated for papers with more than nine recommendations, since the number of papers with less than n=30 were too low (see Table 1). The results are shown in Table 3. Besides the individual coefficients, the tables include the corresponding confidence intervals (Reichenheim, 2004; StataCorp., 2013). Apart from the k coefficient which was calculated for nine Faculty members (k=.25), all ICC and k indicate a low level of agreement between the members – independent of the number of members doing the assessment.

Table 3.

Inter-rater reliability for different numbers of Faculty members

| | Inter-rater reliability | 95% Confidence interval | |
|---|---|---|---|
| **Two members** (n=14,476) | | | |
| Kappa | .14 | .13 | .15 |
| ICC | .21 | .20 | .23 |
| **Three members** (n=4,225) | | | |
| Kappa | .14 | .12 | .17 |
| ICC | .21 | .19 | .23 |
| **Four members** (n=1,602) | | | |
| Kappa | .12 | .07 | .15 |
| ICC | .19 | .17 | .22 |
| **Five members** (n=637) | | | |
| Kappa | .09 | .00 | .13 |
| ICC | .15 | .12 | .18 |
| **Six members** (n=266) | | | |
| Kappa | .06 | -.04 | .13 |
| ICC | .18 | .13 | .23 |
| **Seven members** (n=138) | | | |
| Kappa | .00 | -.13 | .10 |
| ICC | .13 | .08 | .20 |
| **Eight members** (n=72) | | | |
| Kappa | .14 | .01 | .31 |
| ICC | .11 | .05 | .20 |
| **Nine members** (n=49) | | | |
| Kappa | .25 | .09 | .54 |
| ICC | .17 | .10 | .28 |

Table 4.

Inter-rater reliability for two Faculty members who agreed in allocating a publication to a specific category

| Category | N | Kappa | 95% Confidence Interval | |
|---|---|---|---|---|
| Confirmation | 223 | .35 | .21 | .47 |
| Controversial | 29 | | | |
| Good for teaching | 2 | | | |
| Hypothesis | 78 | .17 | -.02 | .34 |
| Negative | 1 | | | |
| New finding | 3,634 | .14 | .11 | .17 |
| Novel drug target | 10 | | | |
| Clinical trial (non-RCT) | 12 | | | |
| Refutation | 5 | | | |
| Review | 14 | | | |
| Systematic review | 26 | | | |
| Technical Advance | 353 | .2 | .09 | .29 |

Note. Kappa coefficients were only calculated for categories with over n=30 publications.

Table 4 shows the inter-rater reliabilities (k) for two Faculty members who both assign a publication to a particular category (e.g. good for teaching). So as to be able to make reliable statements about the reliability, the coefficients are only calculated for categories with more than n=30 publications. When two Faculty members jointly assign a publication to a category, one might expect that they also agree about the quality of the publication in their recommendation. But as the results in the tables show, the values of the coefficients hardly differ from those in Table 3. Three categories show a slight agreement with k = .14, .17 and .2. The k of .35 for the confirmation category can be regarded as fair agreement.

## 4.2 Assessments of validity

Following recommendations, such as those of Harnad (2008) that “peer review … [has] to be evaluated objectively (i.e., via metrics)” (p. 103), the most important step in the assessment of the predictive validity of a certain journal peer review process consists of gauging the impact of the accepted and rejected, but otherwise published manuscripts. As the number of citations to a publication reflects its international impact, and given the lack of other operationalisable indicators, it is common in peer review research to evaluate the success of the process on the basis of citation counts. Citation counts are attractive raw data for the evaluation of research output: They are “unobtrusive measures that do not require the cooperation of a respondent and do not themselves contaminate the response (i.e., they are non-reactive)” (Smith, 1981, p. 84). Although citations have been a controversial measure of both quality and scientific progress (e.g., scholars might cite because the cited source corroborated their own views or preferred methods, rather than because of the significance and relevance of the works cited), they are still accepted as a measure of scientific impact and thus as a partial aspect of scientific quality (Martin & Irvine, 1983). The few studies that have examined the predictive validity of journal peer review on the basis of citation impact indicators confirm that peer review represents a quality filter and works as an instrument for the self-regulation of science (Bornmann, 2011).

Since this study is concerned with the evaluation of a post-publication peer review system, bibliometric indicators can be extracted for all papers included in the F1000Prime database (see section 3.1). In the evaluation of the peer review for a particular journal, the fate of the rejected contributions must be investigated beforehand. Whereas the evaluation of journal peer review is concerned with checking the predictive validity of the recommendations of peers and editors' decisions, in this study the convergent validity of the recommendations of the Faculty members is investigated (Thorngate, Dawes, & Foddy, 2009). Since the publications assessed by the Faculty members are published, the recommendations are included in the F1000Prime database in the time in which the publications are already cited. With predictive validity assessment, an evaluation is made first, then the contribution is published and perhaps cited. A successful evaluation of the convergent validity of F1000Prime recommendations would show that the one indicator for scientific quality (here: recommendations of Faculty members) is highly correlated with another indicator (here: citation impact). Even if citations only measure a partial aspect of quality (the impact of research), a high correlation would indicate that both instruments measure theoretically similar concepts. Thus, high correlations would be evidence of convergent validity.

Table 5.

Number of $P_{top\ 10\%}$ per recommendation score (in percent, the assumption of independent observations is violated by including more than one recommendation scores per paper)

| $P_{top\ 10\%}$ | Recommendation score | | | Total (n=110,341) |
|---|---|---|---|---|
| | 1 (n=63,826) | 2 (n=38,721) | 3 (n=7,794) | |
| 0 | 58.7 | 42.9 | 30.0 | 51.1 |
| 1 | 41.3 | 57.1 | 70.0 | 48.9 |
| Total | 100 | 100 | 100 | 100 |

To identify citation impact differences between the recommendation scores (1, 2, 3), a series of logistic regression models have been estimated (Hardin & Hilbe, 2012; Hosmer & Lemeshow, 2000; Mitchell, 2012). Such models are appropriate for the analysis of dichotomous (or binary) responses. Dichotomous responses arise when the outcome is the presence or absence of an event (Rabe-Hesketh & Everitt, 2004). $P_{top\ 10\%}$ is a binary variable with the value "1" if a paper belongs to the 10% most frequently cited papers within its subject category and publication year and the value "0" otherwise (see Table 5). The violation of the assumption of independent observations by including more than one recommendation scores per paper in the regression model is considered by using the cluster option in Stata (StataCorp., 2013). This option specifies that the scores are independent across papers but are not necessarily independent within the same paper (Hosmer & Lemeshow, 2000, section 8.3).

Adjusted predictions are used to make the results easy to understand and interpret. To get a practical feel for the performance differences of papers with different recommendation scores, the predicted probabilities of $P_{top\ 10\%}$ for the publications with different scores are calculated in a logistic regression model. Such predictions are referred to as margins, predictive margins, or adjusted predictions (Bornmann & Williams, 2013; Williams, 2012; Williams & Bornmann, in preparation). The predictions allow the significance of the empirical results to be determined by the statistical significance test. Even if the F1000Prime data are divided into two samples for checking the convergent validity, the two samples of over 50,000 records are so large that significant results would be expected in a statistical test.[6]

[6] To maintain a statistically significant difference between the two $P_{top\ 10\%}$ proportions of 40% and 50%, for a significance level of 5% and a power of 80% one would only need a sample of around 400 records per group (as with e.g. recommendation score 1 and recommendation score 2).

Table 6.

Logistic regression models for $P_{top\ 10\%}$ as dependent and recommendation scores and JIFs as independent variables

| | (1)<br>Baseline model, sample 1 | (2)<br>Baseline model, sample 2 | (3)<br>Model with JIF included, sample 1 | (4)<br>Model with JIF included, sample 2 | (5)<br>Model with JIF (squared) included, sample 1 | (6)<br>Model with JIF (squared) included, sample 2 |
|---|---|---|---|---|---|---|
| Recommendation score | | | | | | |
| Score 1 (Reference category) | | | | | | |
| Score 2 | 0.688*** | 0.674*** | 0.366*** | 0.336*** | 0.341*** | 0.318*** |
| | (33.59) | (33.00) | (16.75) | (15.35) | (15.49) | (14.45) |
| Score 3 | 1.347*** | 1.223*** | 0.570*** | 0.472*** | 0.584*** | 0.486*** |
| | (29.35) | (27.85) | (11.39) | (9.89) | (11.81) | (10.27) |
| JIF | | | 0.0737*** | 0.0732*** | 0.160*** | 0.153*** |
| | | | (50.10) | (51.02) | (37.37) | (37.39) |
| JIF squared | | | | | -.002*** | -.002*** |
| | | | | | (-21.21) | (-.20.94) |
| Constant | -0.290*** | -0.276*** | -1.075*** | -1.052*** | -1.548*** | -1.498*** |
| | (-22.76) | (-21.64) | (-56.31) | (-55.67) | (-53.2) | (-52.41) |
| N | 55,589 | 55,945 | 55,067 | 55,435 | 47,613 | 47,712 |
| pseudo $R^2$ | 0.028 | 0.024 | 0.106 | 0.102 | 0.116 | 0.111 |

Notes.
t statistics in parentheses
* $p < 0.05$, ** $p < 0.01$, *** $p < 0.001$

Table 6 shows the results for the baseline regression models (samples 1 and 2) which includes the recommendation scores as independent and $P_{top\ 10\%}$ as dependent variables. As the results show, score 2 and score 3 were statistically significantly more often applied to highly cited publications than score 1 (the reference category in the model). Whereas the logistic regression models illustrate which effects are statistically significant and what the direction of the effects is, adjusted predictions can provide us a practical feel for the substantive significance of the findings. Figure 1 shows the adjusted predictions (APs) for the three recommendation scores in the logistic regression model. The figure is helpful in clarifying the magnitudes of the effects of the different scores. The APs for the baseline models (samples 1 and 2) show that about 40% of publications with a score of 1 are highly cited, compared to about 60% of publications with a score of 2 and about 73% of publications with a score of 3.

| **Baseline model** | |
|---|---|
| Sample 1 | Sample 2 |
| 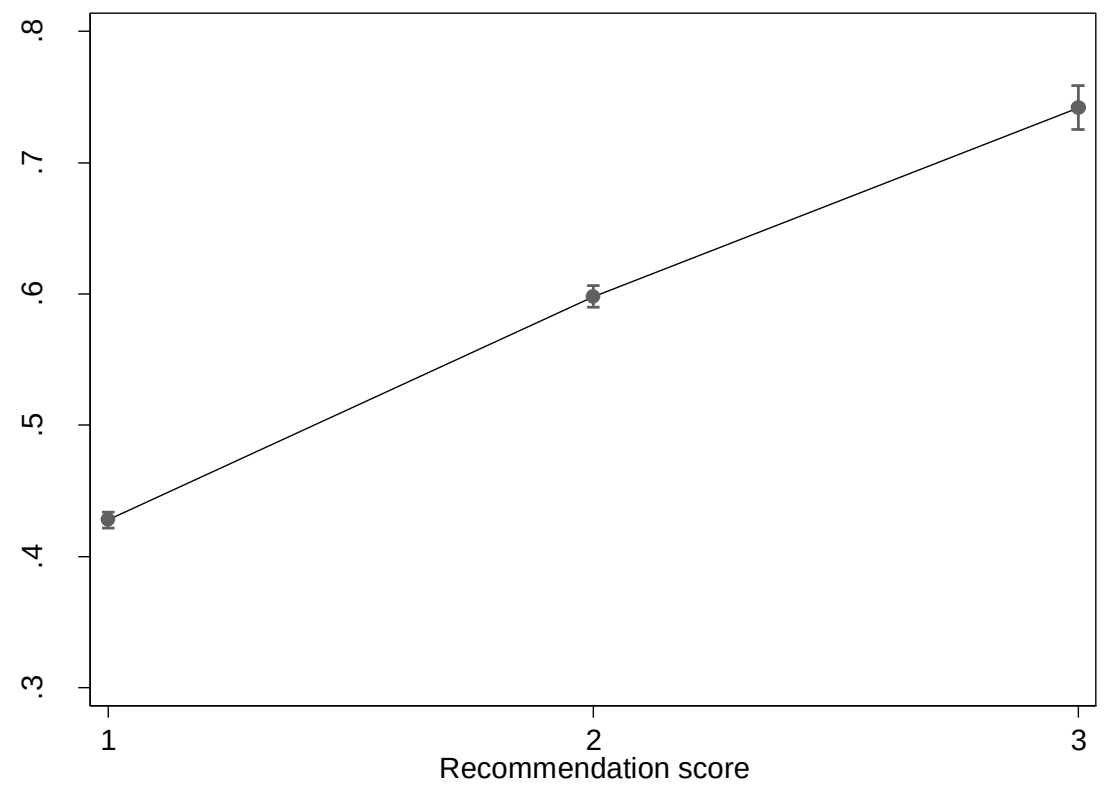 | 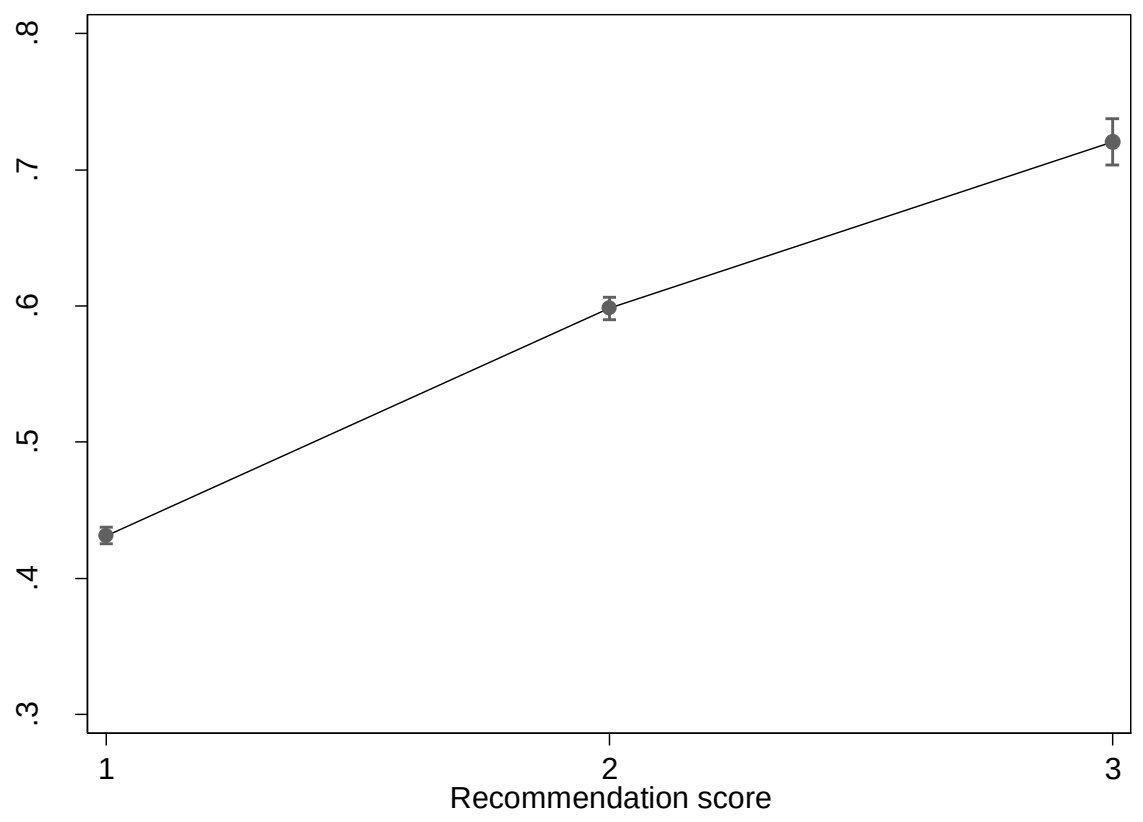 |
| **Taking account of JIF in the model** | |
| Sample 1 | Sample 2 |

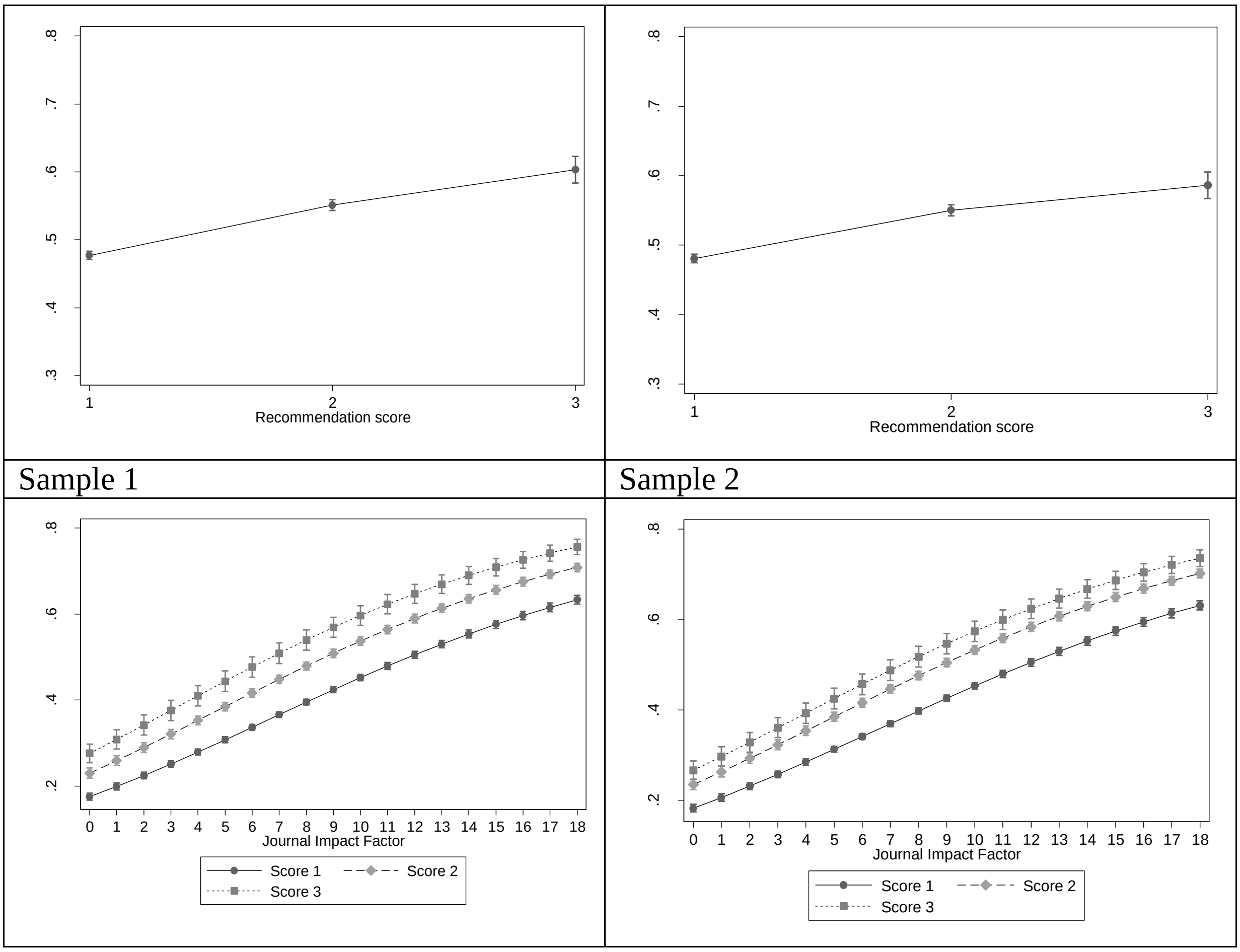

Figure 1. Adjusted predictions (APs and APRVs) and 95% confidence intervals for three recommendation scores and JIFs

This result from the F1000Prime peer review system shows, in agreement with most other results on journal peer review, that a higher citation impact of papers is to be expected with better recommendations from peers – see the overview of results from Bornmann (2011) or the specific results from Buela-Casal and Zych (2010) on the connection between number of citations and the quality evaluated by experts in psychology journals. The result from this study is also in agreement with the results of the previous studies reporting coefficients for the correlation between F1000Prime recommendations and citations. Figure 2 shows the results of a meta-analysis (Glass, 1976) which is based on the correlation coefficients reported by Bornmann and Leydesdorff (2013), Li and Thelwall (2012), Mohammadi and Thelwall (2013) and Waltman and Costas (2014).[7] The pooled correlation coefficient is r=0.246 which can be approximately interpreted as a medium effect size (Cohen, 1988).

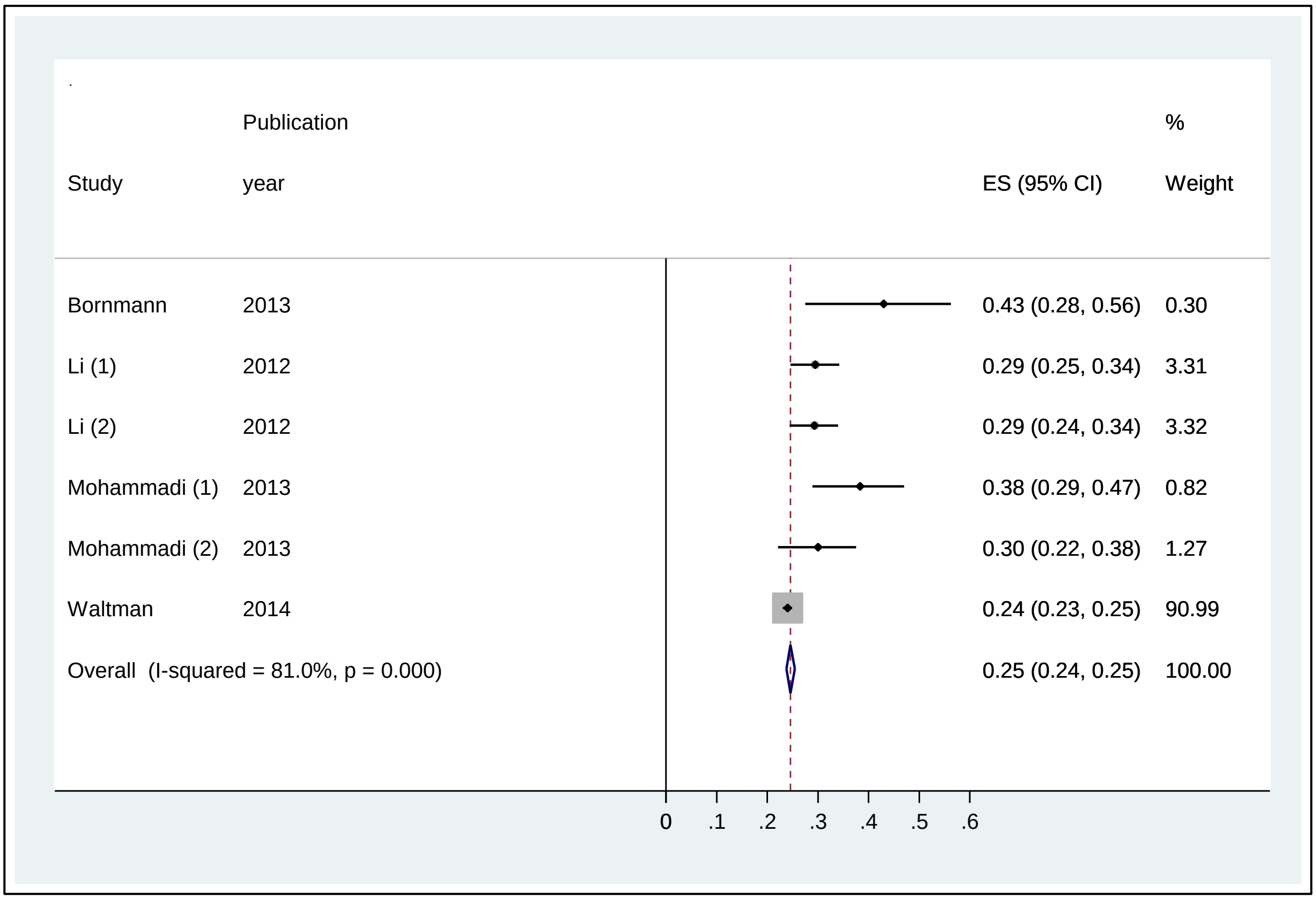

Figure 2. Correlation between F1000Prime recommendations and citations (pooled r=.246). Since Li and Thelwall (2012) report correlation coefficients for WoS and Scopus and Mohammadi and Thelwall (2013) for two years (2007 and 2008), both studies are doubly present. Whereas the study of Waltman and Costas (2014) is based on the maximum F1000Prime recommendation scores, the other studies included the FFa. The studies of Bornmann and Leydesdorff (2013), Li and Thelwall (2012) (1) and Waltman and Costas (2014) used the WoS as data source; the other studies used Scopus. With the exception of Waltman and Costas (2014) who report Pearson correlation coefficients, the studies report Spearman correlation coefficients.

[7] The study of the Medical Research Council (2009) could not be included, because correlation coefficients are not reported.

The advantage of the bibliometric indicator used in this study ($PP_{top\ 10\%}$) is – compared to the use of raw citation counts – that an expected value is available for the analysis: With a publication set we can expect a $PP_{top\ 10\%}$ of 10%. So 10% of the papers should belong to the $P_{top\ 10\%}$ in their publication year and subject area. Since the papers which were given a recommendation of "good" (score 1) by the Faculty members have a share of around 40% in the highly cited papers, the papers in this lowest assessed group are represented significantly more often among the $P_{top\ 10\%}$ (around four times as often) as one would expect. For comparison: In the Leiden Ranking, none of the best institutions worldwide reached a value of over 30% in the $PP_{top\ 10\%}$, (Bornmann & de Moya Anegón, in press). With $PP_{top\ 10\%}$ values of around 60% ("very good", score 2) and around 73% ("exceptional", score 3) highly cited papers, the papers in these two groups reach exceptionally high values. These results agree to the results of the Medical Research Council (2009) that "those papers chosen for evaluation by faculty members do subsequently accumulate a high citation impact."

According to Bornmann, Mutz, Marx, et al. (2011) a paper is more likely to be cited if it is published in a reputable journal rather than in a journal with a poor reputation (see also Lozano, Larivière, & Gingras, 2012; van Raan, 2012). Indeed, if papers receive more citations because they appear in higher-prestige journals and journals gain prestige because they publish papers that receive more citations, then the feedback conditions for self-fulfilling prophecy are in place (Starbuck, 2005). This assertion coincides with the intuition of Garfield (1991), who believes that "the extent of a paper's "citedness" (.. .) is fairly predictable. If it's published in a high-impact journal, it is highly likely to be cited. If it's published in a lower-impact periodical, it may remain uncited – even if it received high marks in prepublication peer review or is frequently read." Thus, the JIF as a measure of the reputation of a journal is included in the regression models of this study. It is interesting to see that the differences between the recommendation scores change substantially with the inclusion of this additional independent variable. Seen overall, the fit of the model is significantly improved by the inclusion of this variable (see the pseudo $R^2$ in Table 6).

While models 3 and 4 fit much better than models 1 and 2, it also makes some questionable assumptions. We might expect diminishing returns for higher JIFs, i.e. it is better to be published in a more influential journal but after a certain point the benefits become smaller and smaller (Williams, 2012). To address such possibilities, in two further models (5 and 6) squared terms for JIF are added. Squared terms allow for the possibility that the variables involved eventually have diminishing benefits or even a negative effect on citation impact (Berry & Feldman, 1985). Since both squared terms are negative, highly significant, and theoretically plausible, models 5 and 6 constitute the final models. Figure 1 shows the APs for the recommendation scores in the logistic regression models 5 and 6 under consideration (control) of the JIF. It is clearly visible that the recommendation scores no longer differ so greatly in the probability of the paper being highly cited. Whereas 48% of the publications with a score of 1 are highly cited, with scores of 2 and 3 it is almost 55% and 60% of the publications.

Figure 1 shows also the adjusted predictions at representative values (APRVs) for the three scores for JIFs ranging between 0 and 18. Extending the JIF range by including larger values than 18 makes the graph hard to read. The graphs for both samples show that, for all three scores, increases in JIFs up to a JIF of around 14 increases the likelihood of the publication being highly cited. With very high JIFs the effect of this indicator on citedness is no longer very clear. Thus with all three scores we can expect a higher probability of the publications being highly cited with increasing JIFs (up to a value of around 14). In addition it becomes clear that the individual scores hardly differ in respect of the relationship between being highly cited and JIF.

# 5 Discussion

Before the background of the requirement for reliability and validity in accordance with the quality criteria for professional evaluations (Bock, 2002) placed on every peer review procedure, this study was concerned with the recommendations formulated in the F1000Prime post-publication peer review system.

Even if – according to Ziman (2000) – intersubjectivity plays a fundamental role in science, the assessments of the reliability of the F1000Prime peer review system show a rather low level of agreement between Faculty members. This result is in agreement with most other studies which have been published on journal peer review (Bornmann, Mutz, & Daniel, 2011). However, in contrast to journal peer review, we cannot (always) assume with F1000Prime peer review that opinions on a paper are arrived at independently (this dependency in the data can lead to distorted ICC or k). Since the recommendations are available in the Internet, Faculty members have access to the recommendations of their colleagues. Against this background one might have expected the recommendations to be more similar. Apparently the Faculty members, even when they see their colleagues' recommendations, reach their judgements in a similarly independent manner as they do in journal peer review. Possibly they feel themselves motivated by the judgements of their colleagues to express another opinion or deal with other aspects than those their colleagues had selected. This last point is described for the journal peer review of the journal *Nature* in an exemplary manner: „In one case, an exciting result relied on two techniques and a theoretical interpretation. The theoretical referee was very positive because the work validated an interesting idea. A specialist in one of the techniques was positive because he could find no flaw in its application. But the third referee uncovered a technical shortcoming in the second technique, and the paper was rejected after the editor assessed the significance of the shortcoming" (Anon, 2006, p. 118).

According to Cole (2000), a low level of agreement among peers reflects the lack of consensus that is prevalent in all scientific disciplines at the "research frontier." Cole (2000) says that usually no one reliably assesses scientific work occurring at the frontiers of research. Since it is very probable that many papers in the F1000Prime database come from the research frontier (a high percentage of papers is highly cited), the missing reliability should come as no surprise. Eckberg (1991) points out that differing judgments in peer review are not necessarily a sign of disagreement about the quality of a paper but may instead reveal differing positions and judgment criteria. In addition, peers tend to be either more critical or more lenient in their judgments (Siegelman, 1991), if they direct their attention to "different points, and may draw different conclusions about 'worth'" (Eckberg, 1991, p. 146). The question of whether the comments of peers are in fact based on different perspectives, positions, and so forth has been examined by only a few empirical studies (Weller, 2002).

This study, in a second analysis step, dealt with the convergent validity of the F1000Prime peer review system. Logistic regression models were used to investigate the relationship between recommendation scores of the Faculty members and the probability that an assessed paper belonged to the top 10% of the most-cited papers ($P_{top\ 10\%}$). As the results show, the proportion of highly cited papers among those selected by the Faculty members is significantly higher than the expected value with $PP_{top\ 10\%}$ – that is 10%. Thus the Faculty members are already selecting papers for F1000Prime for which a performance well above average is to be expected. In addition the results show that better recommendation scores are also connected with better performance of the papers. This result confirms the convergent validity of the recommendation scores and is in agreement with the other studies on the F1000Prime peer review system (Jennings, 2006; Li & Thelwall, 2012; Mohammadi & Thelwall, 2013; Wardle, 2010). Thus Waltman and Costas (2014), for example, write "there turns out to be a clear correlation between F1000Prime recommendations and citations." According to Allen, Jones, Dolby, Lynn, and Walport (2009) "at an aggregate level, after 3 years, there was a strong positive association between expert assessment and impact as measured by number of citations and F1000Prime rating."

Just as in the reliability assessment of the F1000Prime peer review, the problem also exists in the assessment of validity that the independent measurement of quality by Faculty members and citations cannot always be assumed. Since citations can appear straight after a paper is published, Faculty members have the possibility of looking at the citation impact of papers in the corresponding literature databases such as WoS or Scopus (Elsevier). But since in this study a percentile based indicator – a so-called advanced bibliometric indicator – is used to measure the citation impact of a paper in comparison with similar papers (see section 3.3), the independence of the quality measurement compared with the measurement of raw citation counts should be largely ensured. Papers may well have received many citation counts; but the relative impact can be significantly lower when compared with the relevant reference set. Most Faculty members will have no access to advanced bibliometric indicators.

In the framework of the validity analysis of the F1000Prime peer review, this study also investigated which citation impact the papers reach when the JIF is taken into account. The expected value for citations is higher for papers in journals with a high JIF. As the results show, the JIF really does have an influence: the differences in citation impact between papers with different recommendation scores are lower than from the analysis without taking the JIF into account. This result is difficult to interpret since the JIF can influence not only the citation impact but also the scores. In addition it is difficult to separate in the analysis whether the citation impact or the favourable score of a paper coming from quality of research or reputation of a journal. Perhaps the best journals also publish the best papers, or the Faculty members are influenced in their assessment by the JIF of the journal where a paper appears. But since it could be shown in this study that the different recommendation scores show a very similar dependency on JIF and the later citation impact – the higher the JIF the more citation impact can be expected (up to a JIF of around 14) – the citation impact, in particular, appears to be dependent of JIF (and not the recommendation scores). If the individual scores did depend on the JIF, then a similar difference between the scores and the probability of being highly cite would not appear on every level of the JIF. Depending on the level of the JIF, greater or lesser differences between the scores would then have been expected.

The use of the JIF leads to a limitation of this study: In contrast to the citation impact indicators $P_{top\ 10\%}$ and $PP_{top\ 10\%}$, the JIF is not normalized in terms of subject category and publication year. Unfortunately, there is no normalized journal impact indicator (e.g., the source normalized impact per paper, SNIP, Waltman, van Eck, van Leeuwen, & Visser, 2013) available in the MPDL in-house database. Since however all papers included in this study are from subject areas (biology and medicine) with high citation densities, results based on normalized impact factors might be not so different from those reported here.

# 6 Conclusions

Overall the present study agrees with most studies on journal peer review in showing a slight agreement of the Faculty members but a convergent validity between recommendation scores in F1000Prime peer review and citation impact. With the statistical analyses of the F1000Prime peer review system, a further study was able to be added to peer review research, based on a comprehensive dataset of around 100,000 papers. de Vries, Marschall, and Stein (2009) regard this study as urgently needed: “While peer review is central to our science, concerns do exist. Despite its importance, it is curious that we have not required the same rigor of study of the peer review process as we do for our science” (p. 275).

## Acknowledgements

I would like to thank Ros Dignon and Iain Hrynaszkiewicz from F1000 for providing me with the F1000Prime data set and for providing feedback on an earlier version of this paper.
The bibliometric data used in this paper are from an in-house database developed and maintained by the Max Planck Digital Library (MPDL, Munich) and derived from the Science Citation Index Expanded (SCI-E), Social Sciences Citation Index (SSCI), Arts and Humanities Citation Index (AHCI) prepared by Thomson Reuters (Scientific) Inc. (TR®), Philadelphia, Pennsylvania, USA: ©Copyright Thomson Reuters (Scientific) 2014. The data were extracted from the database in February 2014. I would like to thank two anonymous reviewers for their valuable feedback to improve the paper.